\documentclass[aps,twocolumn,superscriptaddress,prl]{revtex4-1}

\usepackage{amsmath,amssymb,amsthm,bm,braket,ascmac,bbm}
\usepackage[pdftex]{graphicx}

\newcommand{\tH}{t_{\rm h}}

\usepackage{color}

\usepackage{ulem}

\usepackage[
colorlinks=true, citecolor=blue, urlcolor=blue, linkcolor=blue,
setpagesize=false, bookmarks=false
]{hyperref}

\usepackage[resetlabels]{multibib}
\newcites{SM}{Refs in Supplemental Material}

\begin{document}


\title{Photoinduced $\eta$-pairing at finite temperatures}

\author{Satoshi Ejima}
\affiliation{Institute of Physics, 
University Greifswald, 17489 Greifswald, Germany}
\affiliation{Computational Condensed Matter Physics Laboratory, 
RIKEN Cluster for Pioneering Research (CPR),
Wako, Saitama 351-0198, Japan}

\author{Tatsuya Kaneko}
\affiliation{Department of Physics, Columbia University, 
New York, NY 10027, USA}

\author{Florian Lange}
\affiliation{Institute of Physics, 
University Greifswald, 17489 Greifswald, Germany}

\author{Seiji Yunoki}
\affiliation{Computational Condensed Matter Physics Laboratory, 
RIKEN Cluster for Pioneering Research (CPR),
Wako, Saitama 351-0198, Japan}
\affiliation{Computational Quantum Matter Research Team, 
RIKEN Center for Emergent Matter Science (CEMS), 
Wako, Saitama 351-0198, Japan}
\affiliation{Computational Materials Science Research Team, 
RIKEN Center for Computational Science (R-CCS), 
Kobe, Hyogo 650-0047, Japan}

\author{Holger Fehske}
\affiliation{Institute of Physics, 
University Greifswald, 17489 Greifswald, Germany}

\date{today}


\begin{abstract}
We numerically prove photoinduced $\eta$-pairing in a half-filled fermionic Hubbard chain at both zero and finite temperature. 
The result, obtained by combining the matrix-product-state based infinite time-evolving block decimation technique and the purification method, applies to the thermodynamic limit. 
Exciting the Mott insulator by a laser electric field docked on via the Peierls phase, we track the time-evolution of the correlated many-body system and determine the optimal parameter set for which the nonlocal part of the $\eta$-pair correlation function becomes dominant during the laser pump at zero and low temperatures. These correlations vanish at higher temperatures and long times after pulse irradiation. In the high laser frequency strong Coulomb coupling regime we observe a remnant enhancement of the Brillouin-zone boundary pair-correlation function also at high temperatures, if the Hubbard interaction is about a multiple of the laser frequency, which can be attributed to an enhanced double occupancy 
in the virtual Floquet state.
\end{abstract}

\maketitle


\textit{Introduction.}-- Optical pumping is not only an excellent tool to investigate complex  
few- and many-body systems but also makes it possible to create new phases of quantum matter with tunable properties~\cite{Ichikawa11,Basov17,PhysRevLett.119.086401,Ishihara19}.  Inducing superconductivity 
by light pulses in low-dimensional materials with strong electronic  correlations is certainly one of the most fascinating 
options in this regard~\cite{Fausti2011,Hu2014,Mitrano2016}. 
Thus, it was not surprising that a whole series of theoretical studies has addressed  
the microscopic modeling  and understanding of this nonequilibrium light-matter-interaction  phenomenon~\cite{SKGetal16,KBRetal16,KWRetal17,PhysRevB.96.045125,PhysRevB.96.064515}.

In this context, the so-called $\eta$-pairing, originally proposed by Yang for the Hubbard model~\cite{Yang89}, has attracted renewed attention~\cite{PhysRevB.94.174503,Kaneko19,PhysRevLett.123.030603,PhysRevB.100.045121,Peronaci2020,Li19,Kaneko19b,Fujiuchi20}. 
Pumping the Mott insulating phase may results in an  excited state with enhanced off-diagonal pair-density-wave correlations, which are 
absent in the ground state~\cite{Kaneko19}.  Here the basic mechanism is the creation of $\eta$-pairs triggered by the nonlinear optical excitation of the system in conjunction with the selection rules. Interestingly, for low-amplitude pulses, the peak structure of the pair correlation function is essentially the same as that obtained for the optical spectrum in the ground state, implying that the photoinduced state might indeed result from an $\eta$-pairing mechanism.

The crucial question is whether these findings will remain valid in the {\it thermodynamic limit} and at {\it finite temperature} $T$. 
Some features, e.g., the stripe structure found in the structure factor of the pair correlations (Fig.~2 of Ref.~\cite{Kaneko19}) by exact diagonalization (ED) of small systems, have been shown to disappear by increasing the system size~\cite{SCES19}, exploiting density-matrix renormalization group (DMRG) and time-evolving block decimation (TEBD) methods~\cite{White92,PhysRevLett.91.147902}. For sure, determining the temporal evolution of an {\it infinite},  {\it driven}, {\it strongly correlated} electron system at $T>0$, is one of the most difficult  problems in solid state theory. Since the fermionic Hubbard model~\cite{Hu63}  can nowadays be realized in optical lattices~\cite{JAKSCH200552,Lewenstein07,Schneider2012,Mazurenko2017}, just as its bosonic counterpart~\cite{GK63}, such a theoretical treatment is indispensable, however, for the interpretation of the experimental data, especially in one spatial dimension.

Despite this difficulty, this Rapid Communication aims at proving the existence of photoinduced $\eta$-pairing in the one-dimensional
half-filled fermionic Hubbard model, directly in the thermodynamic limit and for finite temperatures. For this we exploit unbiased numerical techniques, specifically the infinite TEBD (iTEBD) technique~\cite{PhysRevLett.98.070201} based on an infinite matrix-product-state (iMPS) representation~\cite{Sch11} in combination with the purification method~\cite{Suzuki85,PhysRevLett.93.207204}, which  enables us to monitor the real-time evolution of  thermal states at finite target temperature, as accessible by optical-lattice experiments.

\textit{Model.}---  
Our starting point is the Hubbard Hamiltonian, 
\begin{eqnarray}
 \hat{H}&=&-\tH \sum_{j,\sigma} 
  \big(
   \hat{c}_{j,\sigma}^\dagger \hat{c}_{j+1,\sigma}^{\phantom{\dagger}}
   +\text{H.c.}
  \big)
  \nonumber \\ 
  &&
  +U\sum_{j}\left(
	     \hat{n}_{j,\uparrow}-\tfrac{1}{2}
	    \right)
            \left(
	     \hat{n}_{j,\downarrow}-\tfrac{1}{2}
	     \right)\,,
  \label{hubbard}
\end{eqnarray}
where $\hat{c}_{j,\sigma}^{\dagger}$ ($\hat{c}_{j,\sigma}^{\phantom{\dagger}}$)
creates (annihilates) a fermion with spin projection $\sigma$ ($=\uparrow,\downarrow$) 
at lattice site $j$, and  $\hat{n}_{j,\sigma}=\hat{c}_{j,\sigma}^{\dagger} \hat{c}_{j,\sigma}^{\phantom{\dagger}}$.
 The first term represents the kinetic energy (with nearest-neighbor particle hopping amplitude $\tH$) that acts against the  Coulomb interaction (parametrized by the on-site Hubbard repulsion $U$), which tends to localize the fermions by  establishing a Mott insulating state at half band-filling.   The Hubbard Hamiltonian~\eqref{hubbard} commutes with the operator $\hat{\eta}^2=\tfrac{1}{2}(\hat{\eta}^+\hat{\eta}^-+\hat{\eta}^-\hat{\eta}^+) +  (\hat{\eta}^z)^2$, 
where $\hat{\eta}^z=\tfrac{1}{2}\sum_{j}  (\hat{n}_{j,\uparrow}+\hat{n}_{j,\downarrow}-1)$,  
$\hat{\eta}^+=\sum_j(-1)^j \hat{\Delta}_j^\dagger$,  $\hat{\eta}^-=(\hat{\eta}^+)^\dagger$, and  $\hat{\Delta}_j^\dagger=\hat{c}_{j,\downarrow}^\dagger\hat{c}_{j,\uparrow}^\dagger $ 
denotes the on-site singlet pair creation operator, see~\cite{Supplementary}.

As demonstrated in Ref.~\cite{Kaneko19}, photoinduced $\eta$-pairing states may appear when an external time-dependent field couples to the hopping term via a Peierls phase~\cite{Peierls1933},  $\tH \hat{c}_{j,\sigma}^\dagger \hat{c}_{j+1,\sigma}^{\phantom{\dagger}}  \rightarrow \tH e^{\mathrm{i}A(t)} \hat{c}_{j,\sigma}^\dagger 
                       \hat{c}_{j+1,\sigma}^{\phantom{\dagger}} $, 
where the vector potential  
\begin{eqnarray}
 A(t)=A_0 e^{-(t-t_0)^2/(2\sigma_{\rm{p}}^2)}\cos\left[\omega_{\rm{p}}(t-t_0)\right]
 \label{pump}
\end{eqnarray}
describes a pump pulse with amplitude $A_0$, frequency  $\omega_{\rm p}$ and
width  $\sigma_{\rm p}$, centered at time $t_0$ ($>0$). As a result the Hamiltonian becomes time-dependent, $\hat{H}\to\hat{H}(t)$, and the initial (equilibrium) ground state  evolves (forward) in time:  $|\psi(0)\rangle\to |\psi(t)\rangle$. Numerically such a time evolution can be treated  in an efficient manner by combining TEBD and  second-order Suzuki--Trotter decomposition methods~\cite{PhysRevLett.91.147902}. Hereafter we use $\tH$ ($\tH^{-1}$) as the unit of energy (time), 
and set the time step $\delta t\cdot \tH=0.01$.

In fact, using the iTEBD technique, we directly examine  the time evolution of the pair correlation function,
\begin{eqnarray}
 P(r, t)= \frac{1}{L}\sum_j\langle\psi(t)|\left(\hat{\Delta}^\dagger_{j+r}\hat{\Delta}_j
          +{\rm h.c.}\right) |\psi(t)\rangle\,,
 \label{Prt-infinite}
\end{eqnarray}
in case that the number of lattice sites $L\to\infty$.  At $r=0$, the pair correlation gives twice the number of double occupancy, i.e., 
$P(0,t)=2n_{\rm d}(t)=(2/L) \sum_j \langle\psi(t)|\hat{n}_{j,\uparrow}\hat{n}_{j,\downarrow}|\psi(t)\rangle$.  Most notably, the Fourier transform $\widetilde{P}(q,t)=\sum_r e^{\mathrm{i}qr}P(r,t)$  shows an enhancement after the pulse irradiation that was believed to be indicative of $\eta$-pairing in finite Hubbard clusters~\cite{Kaneko19}.  Since we are particularly interested in longer-range pair correlations, we will also analyze the modified  structure factor  $\widetilde{P}_{r>0}(q,t)=\sum_{r>0} e^{\mathrm{i}qr}P(r,t)$, in which the contribution of the double occupancy $n_{\rm d}(t)$ is excluded.  Let us point out that $\widetilde{P}(q,t)$ obtained by iTEBD in iMPS representation fulfils the relation $\widetilde{P}(\pi,t)=2\langle\psi(t)|\hat{\eta}^+\hat{\eta}^-|\psi(t)\rangle/L$,
which is not the case in any (finite-system) TEBD  calculation with open boundary conditions (OBC), 
see~Ref.~\cite{SCES19} and the Supplemental Material~\cite{Supplementary}.\\


\begin{figure}[tb]
 \includegraphics[width=\columnwidth]{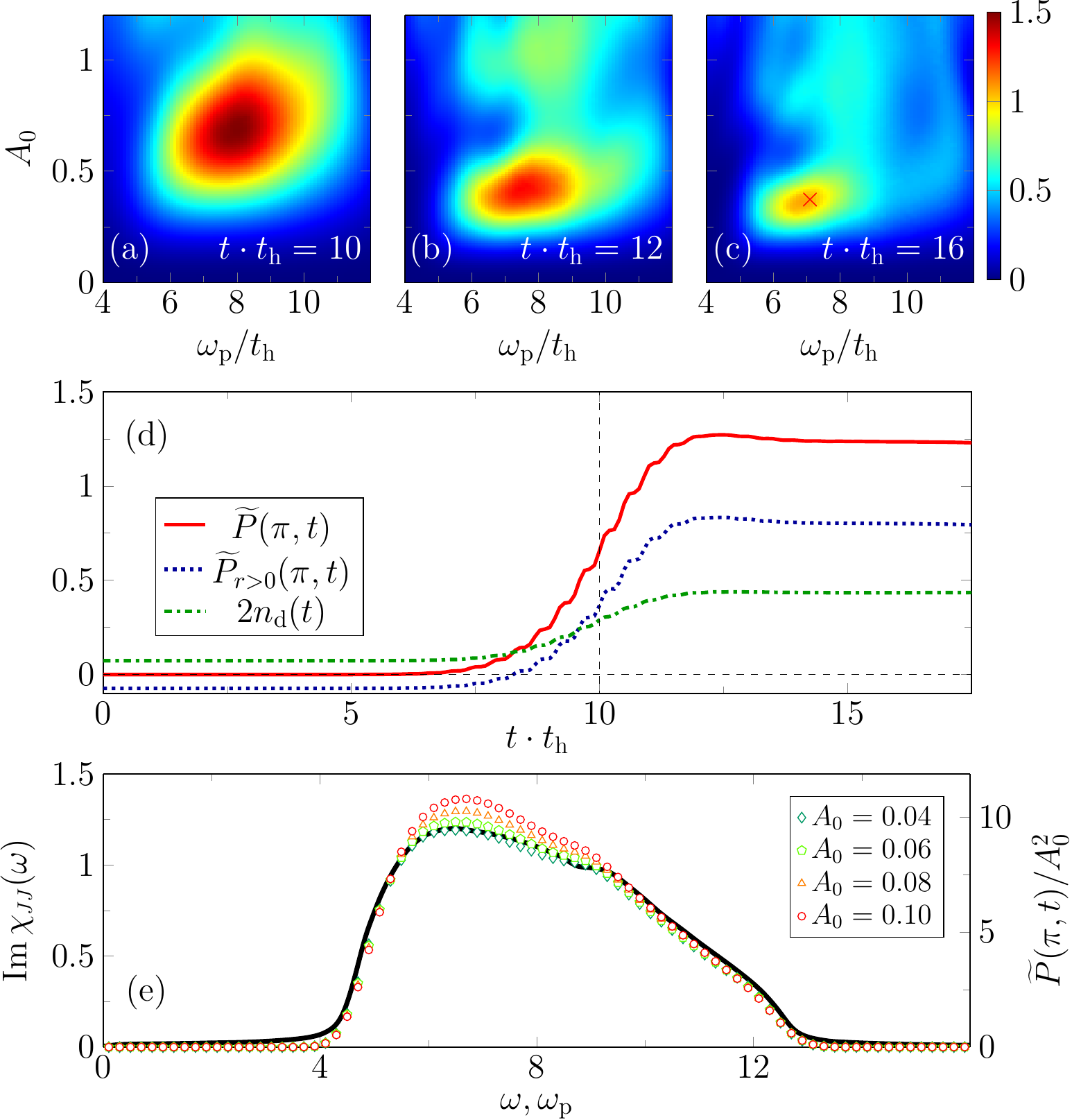}
 \caption{(Color online) Time evolution of pair correlations in an infinite half-filled Hubbard chain at $T=0$. 
 Contour plots of $\widetilde{P}(\pi, t)$ are given in the $\omega_{\rm p}$-$A_0$ plane 
 at $t\cdot \tH=10$ (a), $12$ (b), and $16$ (c) for $U/\tH=8$, where the pump is parametrized 
 by $\sigma_{\rm p}=2$ at $t_0\cdot\tH=10$. $\widetilde{P}(\pi, t)$, $\widetilde{P}_{r>0}(\pi, t)$
 and $2n_{\rm d}(t)$ are displayed as functions of time in (d) for the peak position $\times$ read off from (c).
 Panel (e) demonstrates that the 
 $\widetilde{P}(\pi, t)/A_0^2$ data at $t\cdot t_{\rm h}=16$ (symbols) can be rescaled to 
 ${\rm Im}\chi(\omega)$ (black line) for small $A_0$, where ${\rm Im}\chi(\omega)$ is 
 the imaginary part of the optical spectrum $\chi_{JJ}(\omega)$.
 ITEBD data were obtained with bond dimensions up to $\chi=2000$.  
 For a discussion of the accuracy of the iTEBD calculations see the Supplemental Material~\cite{Supplementary}.
}
 \label{contourT0p3}
\end{figure}
\textit{ITEBD results at $T=0$.}---
In a first step, we determine the optimal parameter set in view of an enhancement of $\widetilde{P}(\pi,t)$
at zero temperature. Figures~\ref{contourT0p3}(a)--(c) provide iTEBD contour plots for 
$\widetilde{P}(\pi, t)$, in dependence on $A_0$ and $\omega_{\rm p}$, at different times  $t\cdot\tH$.
 For $t<t_0$, in the ramp-up regime of the pump field, the spectral intensity of $\widetilde{P}(\pi,t)$ is negligibly small (not shown). 
Noticeable pair correlations develop for $t\gtrsim t_0$, albeit the signal is very broad [cf., Fig.~\ref{contourT0p3}(a)]. 
It becomes focused when the light pulse acts on the system [Fig.~\ref{contourT0p3}(b)], 
and reaches its saturation value
for $t\cdot\tH \simeq 16$ [Fig.~\ref{contourT0p3}(c)], where $A_0=0.37$ and $\omega_{\rm p}=7.10$. 

Figure~\ref{contourT0p3}(d) relates the time evolution of $\widetilde{P}(\pi,t)$,
$\widetilde{P}_{r>0}(\pi,t)$ and $2n_{\rm d}(t)$ for the optimal parameter set 
marked by a cross in Fig.~\ref{contourT0p3}(c). All quantities show a clear response to pulse irradiation and will be strengthened as the system progresses in time until saturation is reached. Apparently, here, the nonlocal contributions $\widetilde{P}_{r>0}(\pi,t)$ have a stronger impact  on $\widetilde{P}(\pi,t)$ than double occupancy.

A notable finding of previous ED calculations~\cite{Kaneko19} was a peak structure of $\widetilde{P}(\pi, t)$ as a function of $\omega_{\rm p}$ which is essentially the same---for small $A_0$---as those of the ground-state optical spectrum $\chi_{JJ}(\omega)$,
folded with an appropriate Lorentzian of width $\eta_{\rm L}$ (depending on $1/\sigma_{\rm p}$). 
The current-current spectral function $\chi_{JJ}(\omega)$ is given by 
\begin{eqnarray}
 \chi_{JJ}(\omega>0)=-\frac{1}{L}\langle\psi_0|\hat{J}\frac{1}{E_0-\hat{H}+\hbar\omega 
                        +\mathrm{i}\eta_{\rm L}}\hat{J}|\psi_0\rangle\,,
 \label{eq-chijj}
\end{eqnarray}
where $|\psi_0\rangle$ is the ground state having energy $E_0$, and the charge current operator $\hat{J}$,  
for the Hubbard model, takes the form $\hat{J}=\mathrm{i}\tH\sum_{\sigma\ell} (
  \hat{c}_{\ell,\sigma}^\dagger \hat{c}_{\ell+1,\sigma}^{\phantom{\dagger}}
 -\hat{c}_{\ell+1,\sigma}^\dagger \hat{c}_{\ell,\sigma}^{\phantom{\dagger}})$. The ED~\cite{Kaneko19} and TEBD~\cite{SCES19}  calculations, which could be conducted for small lattices only, suffer from finite-size effects however. These give rise, inter alia, to stripe patterns in $\widetilde{P}(\pi,t)$, which makes it difficult to determine its maximum value. We demonstrate that a  single peak structure evolves in the thermodynamic limit $L\to\infty$ [see Fig.~\ref{contourT0p3}(c)], and therefore can address more seriously the  question whether the $\chi_{JJ}(\omega)$ lineshape obtained by time-dependent iMPS-based DMRG really agrees with that of $\widetilde{P}(\pi, t)$ for small $A_0$ and large $t$, 
where $\widetilde{P}(\pi)$ becomes time-independent.

Figure~\ref{contourT0p3}(e) compares the iTEBD data, obtained for $\widetilde{P}(\pi,t)$  at various small $A_0$ and $t\cdot\tH=16$,  
with the DMRG results for  $\chi_{JJ}(\omega)$ (using $\eta_{\rm L}/\tH=0.2$), in dependence on $\omega_{\rm p}$ respectively $\omega$. 
Here we show that $\widetilde{P}(\pi, t)$ divided by $A_0^2$ scales to the imaginary part of the optical spectrum 
${\rm Im}\chi(\omega)$ [$\simeq\widetilde{P}(\pi, t)/CA_0^2$ with $C\sim7.9$] since the double occupancy $n_{\rm d}$ is proportional to $A_0^2$, for very small $A_0^2$, in a wide $\omega_{\rm p}$-range around the resonant frequency $\omega_{\rm p}\simeq U$~\cite{PhysRevB.86.075148}.
Close to the maximum in $\widetilde{P}$ respectively  ${\rm Im}\chi$, at about $\omega\simeq6.49$, both quantities differ for larger amplitudes $A_0$, because the nonlocal correlations contained in $\widetilde{P}_{r>0}(\pi,t)$ become increasingly important.
Taking the relation ${\rm Im}\chi_{JJ}(\omega)=\omega\sigma_1(\omega)$ into account, where $\sigma_1(\omega)$ is the real part of the optical conductivity, this behavior is in accordance with DMRG and field-theory results for the optical response in the half-filled Hubbard model~\cite{PhysRevLett.85.3910}.


\begin{figure}[b]
 \includegraphics[width=0.95\columnwidth]{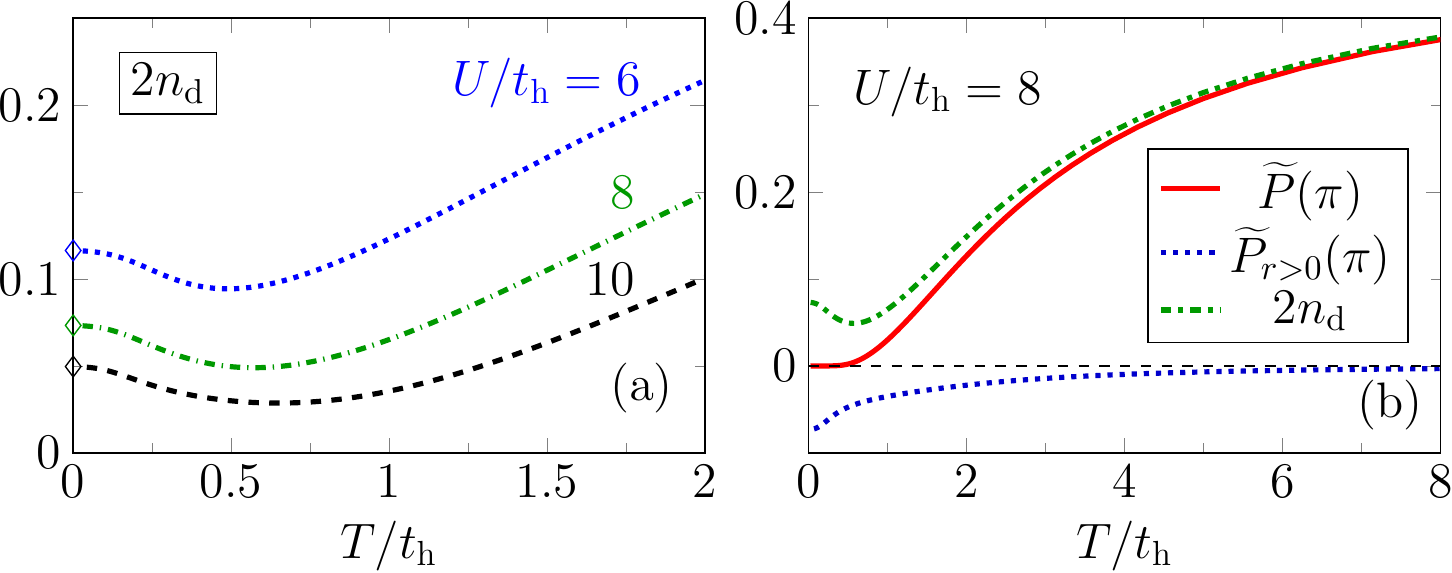}
 \caption{(Color online) Temperature dependence of various correlation functions for the half-filled Hubbard chain 
 without irradiation. Double occupancy $n_{\rm d}$ at different Coulomb repulsions  $U/t_{\rm h}$ (a) [here, the symbols mark data obtained by a separate ground-state simulation ($T=0$)]  and pair correlators $\widetilde{P}(\pi)$, $\widetilde{P}_{r>0}(\pi)$ compared to $n_{\rm d}$
 for $U/t_{\rm h}=8$ (b).
 }
 \label{nud-finT}
\end{figure}
\textit{ITEBD results for $T>0$.}---
In a second step, we  will investigate--under usage of the iMPS and purification approaches~\cite{Suzuki85,PhysRevLett.93.207204}--what happened to the $\eta$-pairing correlations at finite temperatures $T=1/\beta$. Methodically, to obtain the equilibrium state $|\psi_T\rangle$ at some target temperature $T$, we first construct an iMPS representation of a state $|\psi_\infty\rangle$ at infinite temperature, where each physical site is in a maximally entangled state with an auxiliary site, and then carry out  the imaginary-time evolution $e^{-\beta \hat{H}/2}|\psi_\infty\rangle$ 
of the physical system. We note that combining the Suzuki--Trotter decomposition with swap gates~\cite{Stoudenmire2010}, such a  time evolution can be effectively implemented for any nearest-neighbor Hamiltonian.

\begin{figure}[t]
 \includegraphics[width=\columnwidth]{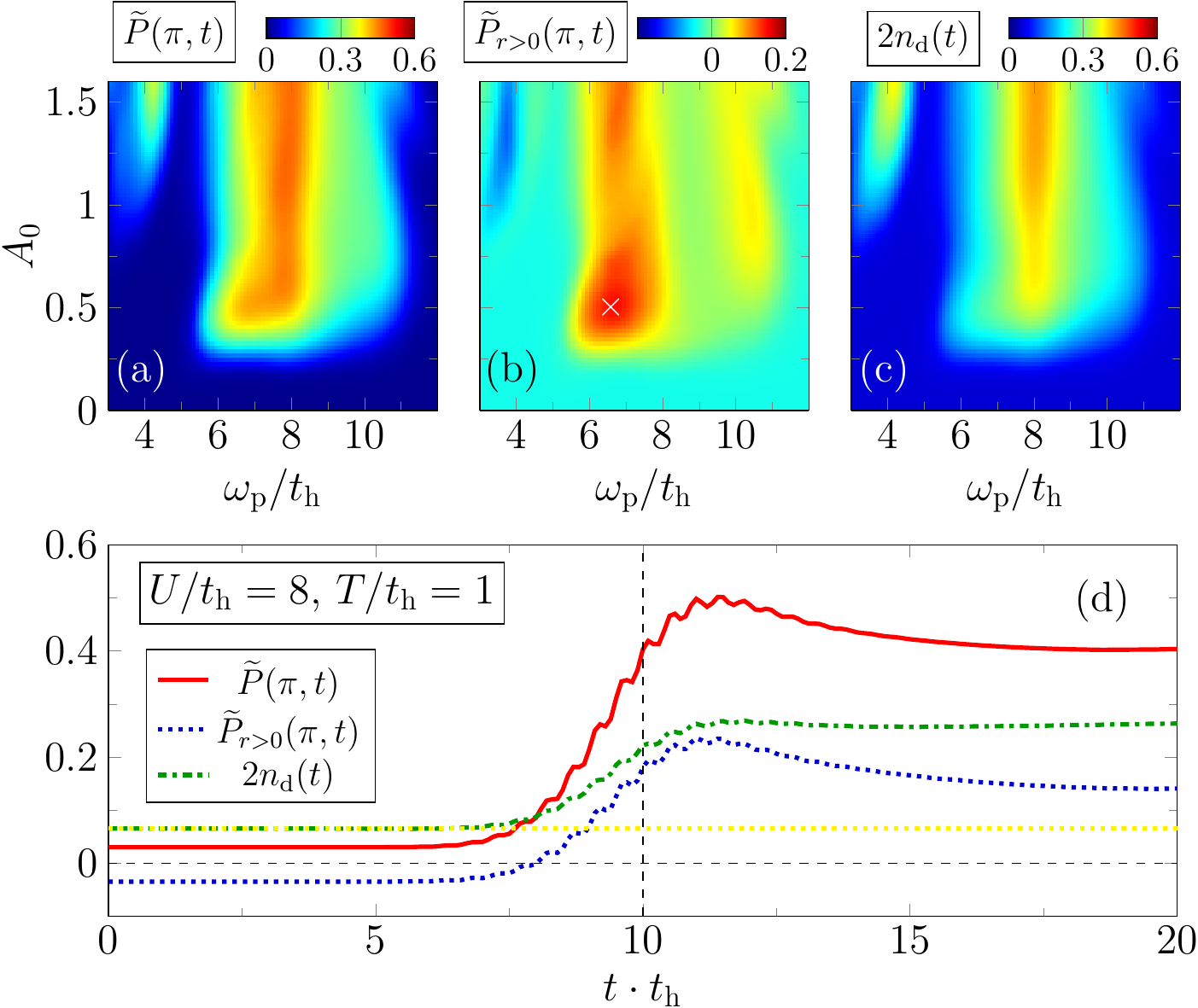}
 \caption{(Color online) Pair correlations in an infinite  half-filled Hubbard chain at $T>0$. Contour plots of $\widetilde{P}(\pi, t)$ (a), $\widetilde{P}_{r>0}(\pi, t)$ (b) and  $2n_{\rm d}(t)$ (c) in the $\omega_{\rm p}$-$A_0$ plane at time $t\cdot\tH=20$ for $T/\tH=1$, after pulse irradiation where $\sigma_{\rm p}=2$, $t_0\cdot\tH=10$, and $U/\tH=8$, obtained by iTEBD with bond dimensions $\chi=800$. 
Time evolution of  $\widetilde{P}(\pi, t)$,  $\widetilde{P}_{r>0}(\pi, t)$ and $2n_{\rm d}(t)$ (d) for parameters corresponding to the peak position $\times$ in (b).  The dotted yellow line marks $2n_{\rm d}(t=0)$ for the pure Hubbard model with corresponding parameters. The iTEBD data are obtained for bond dimension  $\chi=1600$.} 
 \label{T1U8}
\end{figure}

We start by checking the temperature dependence of the double occupancy $n_{\rm d}$ 
in the pure Hubbard model~\eqref{hubbard} without optical pump 

Our iTEBD data in Fig.~\ref{nud-finT}(a) reveal the well-known minimum in $n_{\rm d}$~\cite{Fehske1984}, which is shifted to higher temperatures as $U$ increases and is related to the maximum in the local magnetic moment $L_0=\tfrac{3}{4}\langle (n_{j,\uparrow}-n_{j,\downarrow})^2\rangle$ [$= \tfrac{3}{4}(1-2n_{\rm d})$ at half filling]. At $T=0$, $L_0$ interpolates between the atomic limit ($U=\infty$) with $L_0=3/4$ since $n_{\rm d}=0$ and the band limit ($U=0$) where $L_0=3/8$, i.e., $n_{\rm d}=1/4$, which is also the value for $T\to\infty$ since empty, spin-up/down and double occupied sites are equally likely.  Figure~\ref{nud-finT}(b) shows the temperature dependence of $\widetilde{P}(\pi)$, together with those  of $\widetilde{P}_{r>0}(\pi)$ and $n_{\rm d}$. At $T=0$, on-site [$n_{\rm d}$] and nonlocal [$\widetilde{P}_{r>0}(\pi)$] contributions cancel each other, so that $\widetilde{P}(\pi)=0$. Clearly the pairing correlations vanish in the opposite $T\to\infty$, expressed by the fact that $\widetilde{P}_{r>0}(\pi)\to 0$ and the $\widetilde{P}(\pi)$-curve tends to $2n_{\rm d}$, see also Ref.~\cite{Kaneko19b}. As a result, strong $\eta$-pair correlations can be expected in the low-temperature region at best.

Now, we take into consideration a time-dependent external field and carry out 
the real-time evolution of $\hat{H}(t)$ to a thermal equilibrium state $|\psi_T\rangle$. 
This allows us to discuss the development of $\eta$-pairing correlations as a function of time at $T>0$.
Figures~\ref{T1U8}(a)--(c) provide iTEBD contour plots of $\widetilde{P}(q=\pi, t)$, 
$\widetilde{P}(q=\pi, t)$ and $2n_{\rm d}(t)$ in the $\omega_{\rm p}$-$A_0$ plane 
for $T/\tH=1.0$, at  $t\cdot\tH=20$, following the pulse exposure. We find a persistent enhancement of $\widetilde{P}(\pi, t)$. 
The crucial question is whether this enhancement can be related to the nonlocal part of the pairing correlation function, or simply stems from the on-site (double occupancy) contribution to  $P(r, t)$. The answer can be read off from the contour plot of $\widetilde{P}_{r>0}(\pi, t)$ [Fig.~\ref{T1U8}(b)], which demonstrates its noticeable contribution. Figure~\ref{T1U8}(c) gives the corresponding values of double occupancy $2n_{\rm d}(t)$. Here we find two maxima at about $\omega_{\rm p}\sim U$ and $2\omega_{\rm p}\sim U$ which can be assigned to resonant driving, i.e., to the existence of a Floquet 
virtual state~\cite{WCMD17}.  How $\widetilde{P}_{r>0}(\pi, t)$ and $2n_{\rm d}(t)$ will influence $\widetilde{P}(q=\pi, t)$ over time can be seen in more detail  in Fig.~\ref{T1U8}(d) for $\omega_{\rm p}/\tH=6.6$ and $A_0=0.5$ [$\times$-position in Fig.~\ref{T1U8}(c)]. 
Apparently, all these quantities are growing when the light pulse acts on the correlated system [around $t_0\cdot\tH$ ($=10$)].
Here the (photoinduced) nonequilibrium physics emerges.  Note that the lineshape of $\widetilde{P}(\pi, t)$  (and especially its decay at larger times) is largely determined by $\widetilde{P}_{r>0}(\pi, t)$.  At $t\cdot\tH\gtrsim 20$ saturation is reached. 
The comparison with the pure Hubbard model results shows the predicted dynamical generation of double occupancy~\cite{HR09,MSGMDE18} after pulse irradiation.  

\begin{figure}[tb]
 \includegraphics[width=0.95\columnwidth]{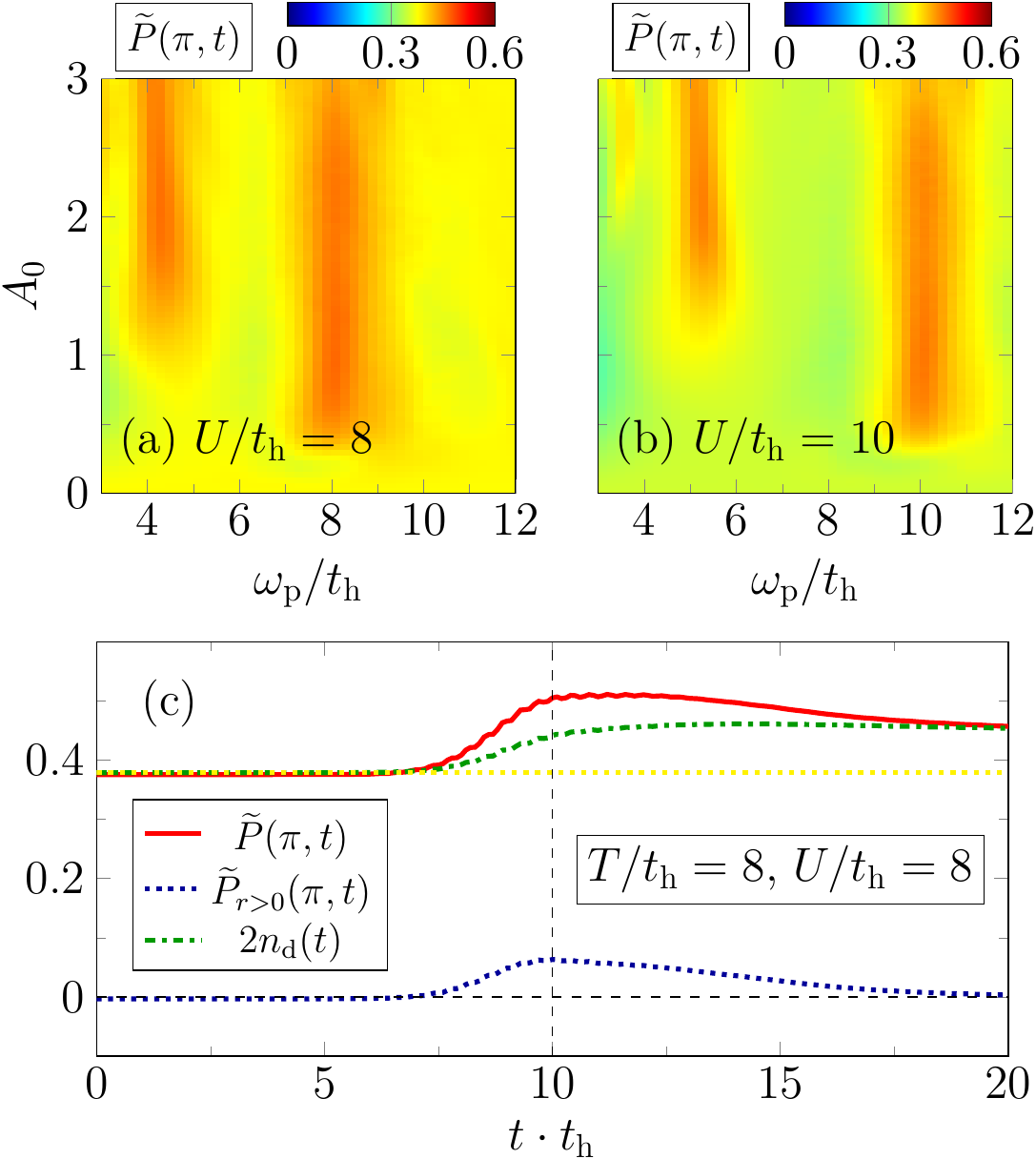}
 \caption{(Color online) Pair correlations at high temperatures. Contour plots of $\widetilde{P}(\pi, t)$ at $T/\tH=8$, calculated for $U/\tH=8$ (a) and $10$ (b) by iTEBD with $\chi=800$, where the pump is parametrized as before. Time evolution of $\widetilde{P}(\pi, t)$,  $\widetilde{P}_{r>0}(\pi, t)$  and  $2n_{\rm d}(t)$ (c), for the peak-position parameters determined from $\widetilde{P}_{r>0}(\pi, t\cdot\tH=15)$ [see Fig.~\ref{pair-corr-T8}(d) in the Supplemental Material~\cite{Supplementary}] by iTEBD with $\chi=1600$. 
 }
 \label{contourU8U10T8}
\end{figure}

Finally, we look at the system response to the pulse at higher temperatures  ($T\sim U$). Figures~\ref{contourU8U10T8}(a) and~(b) display the contour plots of $\widetilde{P}(\pi,t)$, after pumping ($t\cdot\tH=20$),  for $U/\tH=8$ and $U/\tH=10$, respectively.
Again we observe pronounced maxima when the pulse frequency is close to $\omega_{\rm p}\simeq U/m$, which comes to light for  $m=1\,,2$ in (a)  and $m=1\,,2\,,3$ in (b). Figure~\ref{contourU8U10T8}(c) elucidates the origin of this multi-peak structure and the significant differences to the behavior at low-temperatures shown in Fig.~\ref{T1U8}(d).  Before pulse irradiation and at long times 
(where $\widetilde{P}(\pi, t)$ reaches its saturation value), 
$\widetilde{P}(\pi, t)$ is completely determined by $2n_{\rm d}(t)$. The pure Hubbard model result is maintained up to  $t\cdot\tH \simeq 7.5$ (cf. the dotted line), which can be considered as the linear response regime~\cite{WCMD17}. The nonequilibrium dynamics is evidenced at intermediate times $7\lesssim t\cdot\tH \lesssim 13$, when the irradiation is strong.  In contrast to low temperatures, the contribution of  $\widetilde{P}_{r>0}(\pi, t)$ is negligible 
after pulse irradiation for $t\cdot\tH\gtrsim 18$.
This shows that the peak structure observed in Figs.~\ref{contourU8U10T8}(a) and~(b) can be attributed to the enhanced double occupancy. The high-frequency expansion in the Floquet picture reveals the underlying mechanism: Performing a Schrieffer-Wolff transformation~\cite{RGF03,BKP16} for a periodically driven Hubbard model in the strong-coupling regime, an effective (Heisenberg) Hamiltonian is obtained (see, e.g., Ref.~\cite{OK19}), containing an effective exchange interaction $J_{\rm eff}$, which diverges at the resonant frequencies $U\simeq m\omega_{\rm p}$~\cite{HMEW17}.  Since time periodicity $\hat{H}(t+\tau)=\hat{H}(t)$ with $\tau=2\pi/\omega_{\rm p}$ is absent in our model~\eqref{hubbard} and~\eqref{pump}, the photoinduced double occupancy appears as a Floquet virtual state as in the nonequilibrium dynamics of pumped Mott insulators~\cite{WCMD17}. This effect can be observed at any temperature, see, e.g., Fig.~\ref{T1U8}(c).

\textit{Conclusions.}--- 
To sum up, we have demonstrated light-pulse photoinduced $\eta$-paring in the one-dimensional
half-filled Hubbard model at both zero and finite temperatures by means of a defacto approximation-free 
numerical approach. For zero temperature, we carved out finite-size effects of previous exact diagonalization studies, but confirmed the basic relation between the pair correlation function and the ground-state optical spectrum for the infinite system. With a view to experiments, also the optimal pulse for an enforcement of $\eta$--pair correlations $\widetilde{P}(\pi, t)$ is determined. For finite but low temperatures, nonlocal pairing correlations $\widetilde{P}_{r>0}(\pi, t)$ were detected within the applied iTEBD-purification scheme. After pulse irradiation a dynamically generation of double occupancy is proved for finite temperatures.   
Overall, our results support a scenario where optical excitation of a Mott insulator may lead to a (nonequilibrium) state with very slowly decaying pairing correlations. If fermionic optical lattices will be cooled to temperatures $T\lesssim J_{\rm ex}=4\tH^2/U<\tH$ in the strong-coupling regime ($U\gg\tH$)~\cite{Mazurenko2017}, our findings should be detected in the laboratory.

\textit{Acknowledgments} ---
The iTEBD simulations were performed using the ITensor library~\cite{ITensor}.
S.E. and F.L. were supported by Deutsche Forschungsgemeinschaft through Projects No.~EJ~7/2-1
and No.~FE~398/10-1, respectively.  
T.K. acknowledges support from the JSPS Overseas Research Fellowship.
S.Y. was supported in part by Grants-in-Aid for Scientific Research from JSPS
(Project No. JP18H01183) of Japan.

\bibliographystyle{apsrev4-1}

\clearpage

\appendix
\renewcommand\thesection{}
\renewcommand{\theequation}{S\arabic{equation}}
\setcounter{equation}{0}
\renewcommand\thefigure{S.\arabic{figure}}
\setcounter{figure}{0}
\renewcommand{\bibnumfmt}[1]{[S#1]}
\renewcommand{\citenumfont}[1]{S#1}

\section*{Supplemental Material}
\section{$\eta$-pairing symmetry of the Hubbard Hamiltonian in case of periodic boundary conditions}
\label{sec:eta-pairing}
For a one-dimensional lattice with $L$ sites and periodic boundary conditions, the fermionic Hubbard model~\eqref{hubbard} has 
two SU(2) symmetries~\citeSM{HubbardBookS}. Besides the obvious spin symmetry a so-called 
$\eta$-paring symmetry emerges if we look at the operators  
 \begin{align}
  \hat{\eta}^{+}&=\sum_j\hat{\eta}_j^{+}
    =\sum_j(-1)^j\hat{c}_{j,\downarrow}^\dagger\hat{c}_{j,\uparrow}^\dagger\,,
  \nonumber \\
  \hat{\eta}^{-}&=\sum_j\hat{\eta}_j^{-}
    =\sum_j(-1)^j \hat{c}_{j,\uparrow}\hat{c}_{j,\downarrow}\,,
  \nonumber \\
   \hat{\eta}^{z}&=\sum_j\hat{\eta}_j^{z}
    =\sum_j \frac{1}{2}(\hat{n}_{j,\uparrow}+\hat{n}_{j,\downarrow}-1)\,,
 \end{align}
which obey the SU(2) commutation rules
\begin{align}
[\hat{\eta}_j^+,\hat{\eta}_j^-]&=2\hat{\eta}_j^z\,,\\
[\hat{\eta}_j^z,\hat{\eta}_j^{\pm}]&=\pm\hat{\eta}_j^{\pm}\,,
 \end{align}
and satisfy the relationships 
\begin{align}
[\hat{H},\hat{\eta}^{+}\hat{\eta}^{-}]=[\hat{H},\hat{\eta}^{z}]=0\,,
\end{align}
meaning that any eigenstate of the Hubbard Hamiltonian $\hat{H}$ is 
also an eigenstate $|\eta,\eta^z\rangle$ of $\hat{\eta}^2$
and $\hat{\eta}^z$ with eigenvalue $\eta(\eta+1)$ and 
$\eta^z$, respectively.

The presence of $\eta$-pairing states in the Hubbard model was first recognized by Yang~\citeSM{Yang89S}.
He showed that the states $|\phi_{N_\eta}\rangle\propto(\hat{\eta}^+)^{N_\eta}|{\rm vac}\rangle$,
with $|{\rm vac}\rangle$ being the vacuum state and $N_{\eta}$ denoting the 
number of $\eta$-pairs, are eigenstates of $\hat{H}$ which possess off-diagonal long-range order. 
Because these states are excited states, the long-range order does not show up in the ground state or thermal states  of $\hat{H}$ as shown in this study and also Ref.~\citeSM{Kaneko19bS}.

\section{Pair correlation function in case of open boundary conditions}
\label{sec:PairCorrOBC}
For  open boundary conditions (OBC), the pair correlation function 
in real space can be expressed as
\begin{eqnarray}
 P_{\rm OBC}(r, t)=\frac{1}{N_{\rm b}}\sum_{j=1}^{N_{\rm b}}
  \langle\psi(t)|[\hat{\Delta}^\dagger_{j+r}\hat{\Delta}_j
                       +{\rm H.c.}]
  |\psi(t)\rangle,\,
  \label{Prt-OBC}
\end{eqnarray}
where the summation extends  to the number of pairs of sites,  $N_{\rm b}=L-r$, with sites separated by $r$  
in an open chain with $L$ sites~\citeSM{Kaneko19S}. According to~\citeSM{SCES19S}, 
the Fourier transform $\widetilde{P}_{\rm OBC}(q,t)=\sum_r e^{\mathrm{i}qr}P_{\rm OBC}(r,t)$, 
also shows the characteristic enhancement after irradiation which, however, will continue  
to grow for $t>t_0$, in contrast to what is found for periodic boundary conditions. 
Most likely this is caused by the definition of the (quasi-)momenta 
in Eq.~\eqref{Prt-OBC}, in particular at the  boundaries. 
Instead of addressing this issue directly, one can use better the relation 
\begin{eqnarray}
 \widetilde{P}(\pi,t)=\frac{2}{L}\langle\psi(t)|\hat{\eta}^+\hat{\eta}^-|\psi(t)\rangle\,,
  \label{Ppit-etaeta}
\end{eqnarray}
which holds for periodic boundary conditions.

\begin{figure}[t]
 \includegraphics[width=0.9\columnwidth]{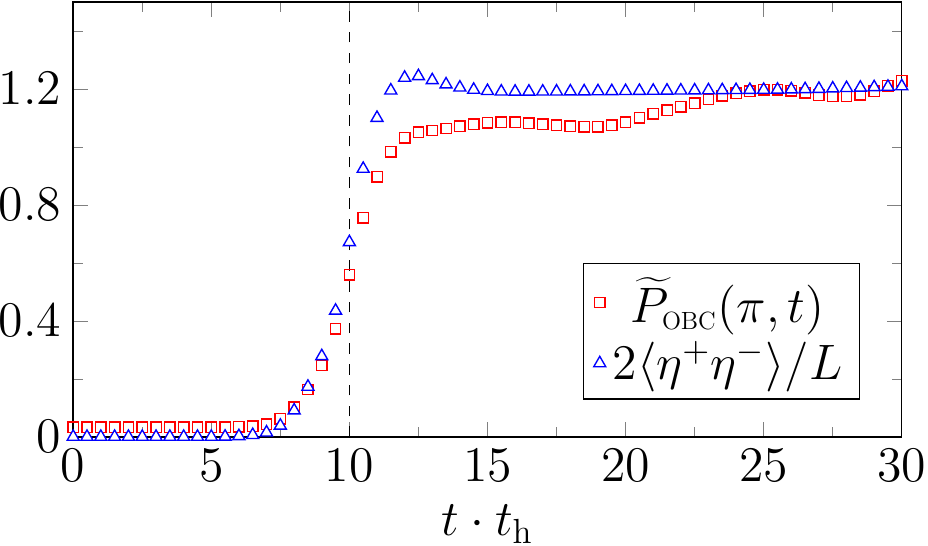}
 \caption{(Color online) Time evolution of the pair correlation function $\widetilde{P}_{\rm OBC}(\pi, t)$ and the  number of $\eta$-pairs $2\langle\hat{\eta}^+ \hat{\eta}^-\rangle/L$ in  an irradiated half-filled Hubbard chain with $U/\tH=8$. Again $\sigma_{\rm p}\cdot\tH=2$ and $t_0\cdot\tH=10$. Now results obtained by TEBD for a finite system with $L=16$  and OBC. 
 }
 \label{ppi-etapetam}
\end{figure}

Figure~\ref{ppi-etapetam} shows the time-evolving block decimation (TEBD) results 
for $\widetilde{P}_{\rm OBC}(\pi, t)$ and 
$\langle\hat{\eta}^+ \hat{\eta}^-\rangle$ in the half-filled
Hubbard model, where we have parametrized the pump by $A_0 = 0.38$ and $\omega_{\rm p}/\tH=6.8$ (this corresponds to the peak position of the $A_0$-$\omega_{\rm p}$ contour plot at $t\cdot\tH=30$, see Fig.~2 of Ref.~\citeSM{SCES19S}.) 
After pulse irradiation, for $t\cdot\tH\gtrsim12$, the magnitude of $2\langle\hat{\eta}^+ \hat{\eta}^-\rangle/L$ saturates 
to a constant value, reflecting the conservation law of the $\eta$-pair numbers,
even though $\widetilde{P}_{\rm OBC}(\pi,t)$ is still weakly growing. 
Note that the relation~\eqref{Ppit-etaeta} will be fulfilled within an infinite TEBD (iTEBD) calculation:
Here, $\widetilde{P}(\pi,t)$  saturates to a constant value, provided the bond dimension ia 
large enough, see Fig.~\ref{contourT0p3}(d) of the main text. \\

\section{Nonlocal pairing correlation $\widetilde{P}_{r>0}(\pi,t)$ at $T/\tH=8$}
\begin{figure}[tb]
 \includegraphics[width=0.9\columnwidth]{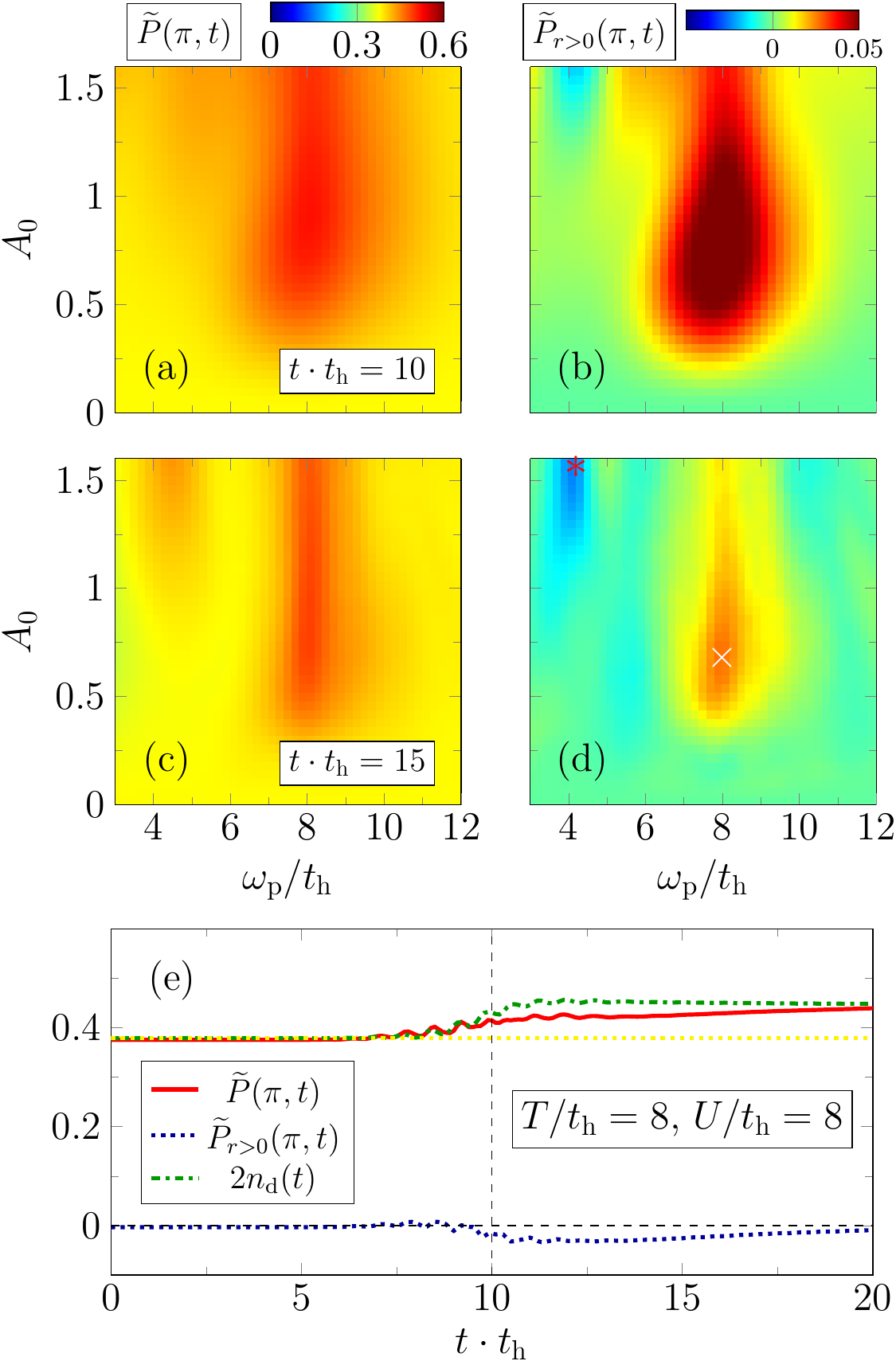}
 \caption{(Color online) Contour plots of $\widetilde{P}(\pi,t)$ [(a) and (c)] and 
 $\widetilde{P}_{r>0}(\pi,t)$ [(b) and (d)]  at $t\cdot\tH=10$ [(a) and (b)]
 and $15$ [(c) and (d)], where $U/\tH=8$ and $T/\tH=8$. The pump is parametrized as 
 $t_0\cdot\tH=10$ and $\sigma_{\rm p}=2$. Data obtained by iTEBD with $\chi=800$.
 Time evolution of $\widetilde{P}(\pi, t)$,  $\widetilde{P}_{r>0}(\pi, t)$  and  $2n_{\rm d}(t)$ (e), for the parameter set marked by the $\ast$ in (d).   
 }
 \label{pair-corr-T8}
\end{figure}
In this section we are presenting further details about the on-site and nonlocal pair correlations 
at high temperatures. Figures~\ref{pair-corr-T8}(a)-(d) give the contour plots
of $\widetilde{P}(\pi,t)$ and $\widetilde{P}_{r>0}(\pi,t)$ for $U/\tH=8$ and $T/\tH=8$.  At $t_0\cdot\tH=10$, 
both pair correlation functions are finite in a large range of parameter space, 
but the enhancement is rather weak  compared to those at zero temperature. 
After pulse irradiation ($t\cdot\tH=15$), the spectral intensity of $\widetilde{P}(\pi,t)$ and $\widetilde{P}_{r>0}(\pi,t)$
becomes concentrated in two spots with driving frequencies $\omega_{\rm p}$  close to the Hartree energy $U/2$ and to the Hubbard interaction energy $U$.   
Interestingly, the peak with $\omega_{\rm p}\simeq U$ ($U/2$) has 
positive (negative) spectral weight in the nonlocal contribution $\widetilde{P}_{r>0}(\pi,t)$
[see Fig.~\ref{pair-corr-T8}(d)]. Note that Fig.~\ref{contourU8U10T8}(c) in the main text
shows $\widetilde{P}(\pi,t)$, $\widetilde{P}_{r>0}(\pi,t)$ and $2n_{\rm d}(t)$
at the peak position of Fig.~\ref{pair-corr-T8}(d) at $\omega_{\rm p}=U=8.0$ and $A_0=0.68$, which is marked by the white cross.  
By contrast, in Fig.~\ref{pair-corr-T8}(e), we show $\widetilde{P}(\pi,t)$, $\widetilde{P}_{r>0}(\pi,t)$ 
and $2n_{\rm d}(t)$ at  $\omega_{\rm p}\approx U/2$, marked by the white asterisk in Fig.~\ref{pair-corr-T8}(d), where $\omega_{\rm p}=4.2$ and $A_0=1.56$. 
As in Fig.~\ref{contourU8U10T8}(c), $\widetilde{P}_{r>0}(\pi,t)$ becomes zero for long times, but now it approaches its limiting value from below.  
Thus  double occupancy $n_{\rm d}(t)$ dominates $\widetilde{P}(\pi,t)$ 
after pumping at high temperatures.

\section{Accuracy of the iTEBD simulations}
\label{sec:accuracy}

\begin{figure}[tb]
 \includegraphics[width=0.9\columnwidth]{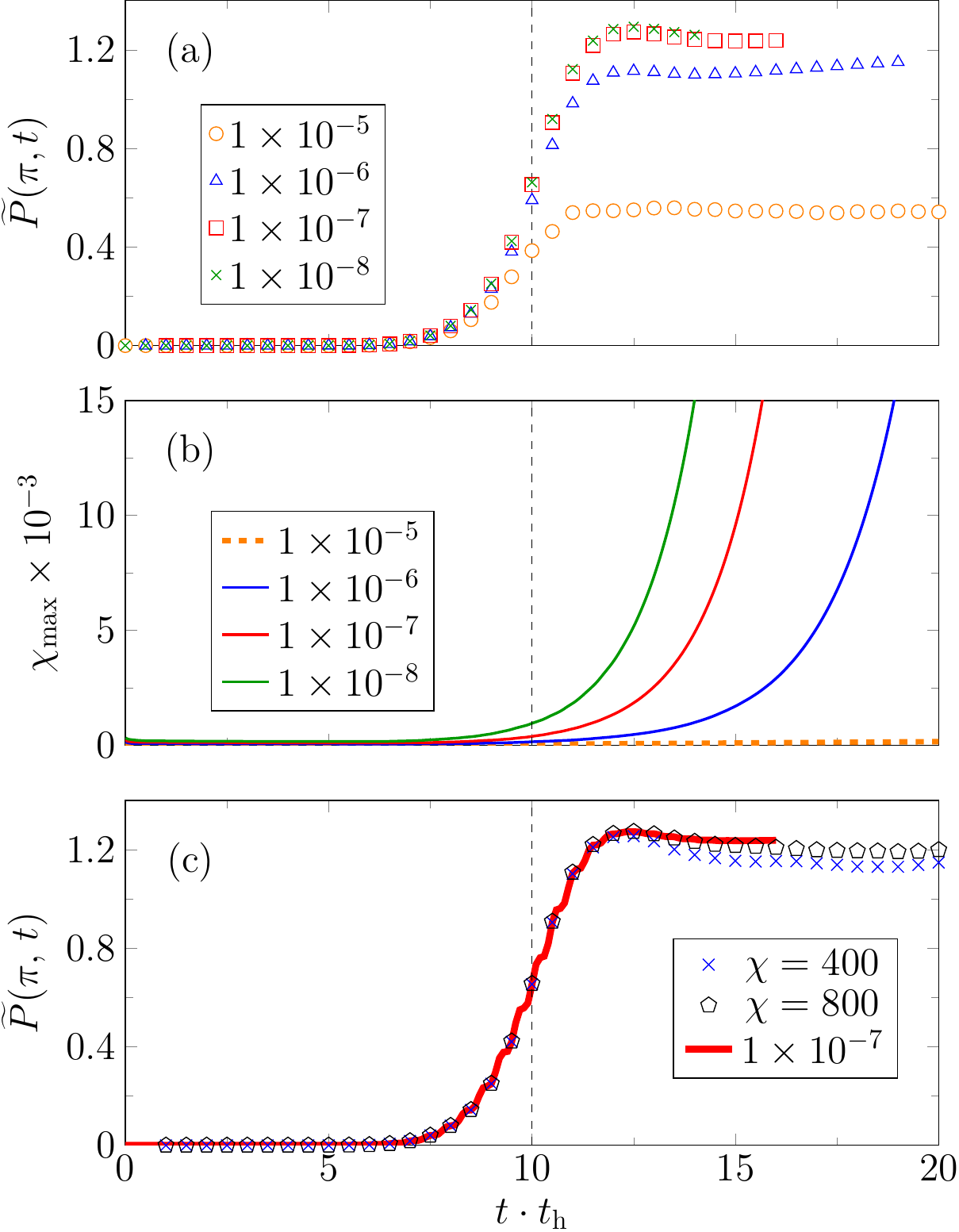}
 \caption{(Color online) (a) Cutoff dependence of $\widetilde{P}(\pi, t)$ for an iTEBD simulation performed at $T=0$, where $U/\tH=8$, $\sigma_{\rm p}=2$ and $t_0\cdot\tH=10$. (b) Maximum bond dimension $\chi_{\rm max}$ needed to 
 keep the  truncation error smaller than a specified value at any time. 
 (c) $\widetilde{P}(\pi, t)$ for various fixed bond dimensions compared with the result given in (a) for a cutoff error less than $10^{-7}$. Here, $A(t)$ is parametrized by $A_0=0.37$ and $\omega_{\rm p}/\tH=7.10$, which are read off from the peak position in Fig.~\ref{contourT0p3}(c) in the main text.
 }
 \label{ppit-maxm}
\end{figure}

In Ref.~\citeSM{SCES19S} we compared the results of TEBD simulations with OBC with exact diagonalization (ED)  data, 
and demonstrated  good agreement up to some time $t\cdot\tH$, depending on the maximum bond dimension $\chi_{\rm max}$ used.   
On the other hand, keeping the cutoff less than $10^{-7}$ during the TEBD simulation
for system size $L=12$ and OBC, a perfect agreement with ED data is also achieved  
for  $t\cdot\tH\lesssim 30$, if $\chi_{\rm max}$ is of the order $10^4$. Since an analytical solution for the time-evolution of  the infinite Hubbard model~\eqref{hubbard} after pulse irradiation  
is lacking, we discuss the accuracy of our iTEBD approach in the latter way. 

Figure~\ref{ppit-maxm} presents iTEBD results for $\widetilde{P}(\pi, t)$
at zero temperature, obtained enforcing various maximum truncation errors.
The discrepancies between the iTEBD data with maximum truncation errors
$1\times10^{-7}$ and $1\times10^{-8}$ are negligible [see panel (a)], albeit the simulations 
have to be performed up to different $t\cdot\tH=16.0$ and $14.4$  because of the rapid 
increase of $\chi_{\rm max}$ needed [see panel (b)]. Performing the iTEBD simulations 
with a cutoff less than $10^{-7}$ may not always be realistic, however, in view of limited computational resources.  
Fortunately, Fig.~\ref{ppit-maxm}(c) demonstrates that a reasonable accuracy can be obtained quite often 
using smaller $\chi_{\rm max}$  (see the results for   $\chi_{\rm max}=400$).\\

\section{Temperature-dependence of the spin structure factor}
\begin{figure}[h]
 \includegraphics[width=\columnwidth]{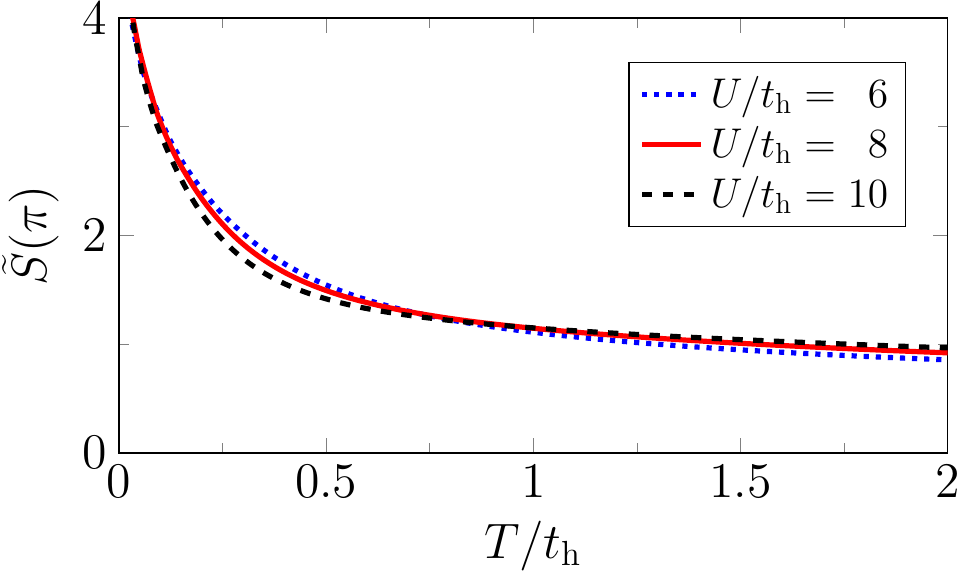}
 \caption{(Color online) Spin structure factor $\widetilde{S}^{zz}(q=\pi)$  as a function of temperature $T$ at various $U$, where $\sigma_{\rm p}=2$ and $t_0\cdot\tH=10$.
 }
 \label{Spi-finT}
\end{figure}

Besides the $\eta$-pair correlations, it is of importance to determine 
the competing antiferromagnetic spin correlations~\citeSM{Kaneko19S}.
In real space, the $zz$-spin correlation function is given by 
\begin{eqnarray}
 S^{zz}(r)= \frac{4}{L}\sum_j\langle\psi|\hat{S}^z_{j+r}\hat{S}^z_{j}|\psi\rangle\,,
 \label{Srt}
\end{eqnarray} 
and $\widetilde{S}^{zz}(q)=\sum_r e^{\mathrm{i}qr}S^{zz}(r)$  is the corresponding spin structure factor in momentum space.
At $q=\pi$, this quantity should monotonously decrease with increasing temperature solely because $\widetilde{S}^{zz}(\pi)\to1/2$ 
for $T\to\infty$. This is confirmed by Fig.~\ref{Spi-finT} showing that $\widetilde{S}^{zz}(\pi)$ is quickly reduced as $T \to \tH$,
and thereafter very slowly approaches its $T=\infty$ limiting value.

%

\end{document}